\documentstyle[12pt]{article}
\title{Leading Quantum Correction to the Newtonian Potential}

\author{John F. Donoghue\\[5mm]
Department of Physics and Astronomy\\
University of
Amherst, MA ~01003}
\date{}
\begin{document}
\begin{titlepage}
\maketitle
\begin{abstract}

I argue that the leading quantum corrections, in powers of the energy
or inverse powers of the
distance, may be computed in quantum gravity through knowledge of only
the low energy
structure of the theory.  As an example, I calculate the leading
quantum corrections to the
Newtonian gravitational potential.
\end{abstract}
{\vfill UMHEP-396}
\end{titlepage}

The Newtonian potential for the gravitational interactions
\begin{equation}
V(r) = - {Gm_1 m_2 \over r}
\end{equation}

\noindent  is of course only approximately valid. For large masses
and/or large velocities there are
relativistic corrections which have been calculated within the framework
of the general theory of
relativity \cite{1}, and which have been verified experimentally.  At
microscopic distance scales,
we would also expect that quantum mechanics would lead to a modification
in the gravitational
potential in much the same way that the radiative corrections of quantum
electrodynamics (QCD)
leads to a modification of the Coulombic interaction \cite{2}. The present
paper addresses these
quantum corrections to the gravitational interaction.

General relativity forms a very rich and subtle classical theory.
However, it has not been possible
to combine general relativity with quantum mechanics to form
a satisfactory theory of
quantum gravity.  One of the problems, among others, is that
general relativity does not fit the
present paradigm for a fundamental theory; that of a renormalizable
quantum field theory.
Although the gravitational fields may be successfully quantized
on smooth-enough background
space-times \cite{3}, the gravitational interactions are of
such a form as to induce divergences
which cannot be absorbed by a renormalization of the parameters
of the minimal general relativity
\cite{3,4,5}.  If one introduces new coupling constants to absorb
the divergences, one is led to an
infinite number of free parameters.  This lack of predictivity
is a classic feature of
nonrenormalizable field theories.  The purpose of this paper is
to argue that, despite this situation,
the leading long distance quantum corrections are reliably calculated
in quantum gravity.  The idea
is relatively simple and will be the focus of this letter, with
more details given in a subsequent paper
\cite{6}.

The key ingredient is that the leading quantum corrections
at large distance are due to the
interactions of massless particles and only involve their
coupling at low energy.  Both of these
features are known from general relativity even if the full
theory of quantum gravity is quite
different at short distances.

The action of gravity is determined by an invariance under general
coordinate transformations, and
will have the form
\begin{equation}
S = \int d^4 x \sqrt{-g} \left[  {2 \over \kappa^2} R + \alpha R^2
+ \beta R_{\mu \nu} R^{\mu \nu} +
\gamma R_{\mu \nu} R^{\nu \alpha} R_\alpha^\mu + \right]
\end{equation}

\noindent  [We ignore the possibility of a cosmological constant,
which experimentally must be
very small].  Here R is the curvature scalar, $R_{\mu \nu}$ is the Ricci
tensor, $g=det g_{\mu
\nu}$ and $g_{\mu \nu}$ is the metric tensor.  Experiment determines \cite{1}
$\kappa^2 = 32 \pi G$,
where G is Newton's constant, and \cite{7} $\mid \alpha \mid, \mid
\beta \mid \leq 10^{74}$.
The minimal general relativity consists of keeping only the first
term, but higher powers of R are
not excluded by any known principle.  The reason that the bounds
on $\alpha, \beta$ are so poor is
that these terms have very little effect at low energies/long
distance.  The quantities R and $R_{\mu
\nu}$ involve two derivatives acting on the
gravitational field (i.e., the metric $g_{\mu \nu}$).  In
an interaction each derivative becomes a factor of the momentum
transfer involved, q, or of the
inverse distance scale $q \sim \hbar / r$.
We will say that R is of order $q^2$.  In contrast, $R^2$
or $R_{\mu \nu} R^{\mu \nu}$ are of order $q^4$.  Thus, at
small enough energies, terms of order
$R^2, R^3$ etc. are negligible and we automatically reduce to only the
minimal theory.

The quantum fluctuations of the gravitational field may be expanded about a
smooth background
metric \cite{3}, which in our case is flat space-time
\begin{eqnarray}
g_{\mu \nu} = \eta_{\mu \nu} + \kappa h_{\mu \nu} \nonumber\\
\eta_{\mu \nu} = diag (1,-1,-1,-1)
\end{eqnarray}

\noindent  About a decade ago, there was extensive study
of the divergences induced in
one and two loops diagrams, also including matter fields \cite{3,4,5,8,9}.
 When starting from the Einstein
action, the divergences appear at higher order, i.e., in $\alpha,
\beta$ for one loop, and in
$\gamma$ at two loops.  This is not hard to see on dimensional
grounds; the expansion is in
powers of $\kappa^2 q^2 \sim Gq^2$ which forms a dimensionless
combination.  These divergences
can be absorbed into renormalized values of the parameters $\alpha,
\beta, \gamma$, which could in principle be determined by experiment. As
mentioned before, higher loops will require
yet more arbitrary parameters.

However, also contained in one loop diagrams are finite corrections
of a different character.  These
are non-analytic contributions, which around flat space
have the form $\kappa^2 q^2 ln(-q^2)$ or $\kappa^2
q^2 \sqrt{{m^2}\over{-q^2}}$.  Because these are non-analytic,
e.q., picking up imaginary parts
for timelike $q^2 (q^2 > 0)$, they cannot be absorbed into a
renormalization of parameters in a
local Lagrangian.  Also, because $\mid ln(-q^2) \mid \gg 1$ and $\mid
\sqrt{{m^2}\over{-q^2}}
\mid \gg 1$ for small enough $q^2$, these terms will dominate over
$\kappa^2 q^2$ effects in the limit
$q^2 \rightarrow 0$.  Massive particles in loop diagrams do not produce
such terms; a particle with
mass will yield a local low energy Lagrangian when it is integrated
out of a theory, yielding contributions to the parameters
$\alpha, \beta, \gamma$ of the Lagrangian in Eq.2 .  In contrast, non-analytic
contributions come from long distance propagation, which at low
energy is only possible for
massless particles.  Similarly, to determine the coefficients of
the long distance non-analytic terms,
one does not have to know the short distance behavior of the theory;
only the lowest energy
coupling are required.  Since both the enumeration of the massless
particles and the low energy
coupling constant follow from the Einstein action, this is sufficient
to determine the dominant low
energy corrections.

The above argument is at the heart of the paradigm of effective
field theories \cite{10,11}, which
have been developed increasingly in the past decade.  Indeed it
is almost identical to the way that
low energy calculations involving pions are performed in chiral
perturbation theory, which is an
effective field theory representing the low energy limit of QCD.  [There the
role of $\kappa^2$ is taken
by $1/16 \pi^2 F_\pi ^2 \approx 1/(1 GeV)^2$ and the higher order
renormalized constants
equivalent to $\alpha, \beta$ are of order $10^{-3}$.]  The interested
reader is directed to the
literature of chiral perturbation theory \cite{10,11,12} to see how an
effective field theory works
in practice, including comparison with experiment.  It has recently
been shown that the sicknesses
of $R + R^2$ gravity are not problems when treated as an effective
field theory \cite{13}.

Let us see how this technique works in the case of the Newtonian
potential.  When one adds a
heavy external source, use of the action of Eq. 2 plus one
graviton exchange leads to a classical potential
of the form \cite{7}
\begin{eqnarray}
V(r) &=& {G m_1 m_2 \over r} \left[ 1 - {4\over 3} e^{- r/r_2} + {1\over 3}
e^{- r/r_0} + \ldots
\right] \nonumber\\
&=& G m_1 m_2 \left[ {1\over r} - 128 \pi^2 G(\alpha + \beta)
\delta^3 (\bf{x} ) + \ldots \right]
\nonumber\\
r_2^2 &=& -16 \pi G \beta \nonumber\\
r_0^2 &=& 32 \pi G (3 \alpha + \beta)
\end{eqnarray}

\noindent  Simply put, the effect of the order $q^4$ effects of $R^2$
and $R_{\mu \nu} R^{\mu
\nu}$ are short ranged. [The second line above indicates that these
terms limit to a Dirac delta
function as $\alpha, \beta \rightarrow 0$. This second form of the potential is
most appropriate for a perturbation in an effective field theory. ]
In contrast the leading quantum
corrections will fall like
powers of r, and hence will be dominant at large r.

In order to calculate the quantum corrections we need to specify the
propagators and vertices of the theory.  It is most convenient to use
the harmonic gauge, $2 \partial_\mu h^\mu _\nu = \partial_\nu
h^\lambda_\lambda$, which is accomplished by including the following
gauge fixing term
\begin{equation}
{\cal L}_{gf} = {\sqrt{-g}} \left[ D_\sigma h^\sigma_\mu -
{1 \over 2} D_\mu h^\sigma_\sigma \right] g^{\mu \nu}
\left[ D_\lambda h^\lambda_\nu - {1 \over 2} D_\nu h^\lambda
_\lambda \right]
\end{equation}

\noindent The most useful feature of this gauge is the relative simplicity
of the  graviton
propagator, which assumes the form
\begin{eqnarray}
D_{\mu \nu, \alpha \beta} (q) &=& {i\over q^2} P_{\mu \nu, \alpha \beta }
\nonumber\\
P_{\mu \nu, \alpha \beta} &=& {1\over 2} \left[ \eta_{\mu \alpha}
\eta_{\nu \beta} + \eta_{\mu
\beta}  \eta_{\nu \alpha} - \eta_{\mu \nu} \eta_{\alpha \beta} \right]
\end{eqnarray}

\noindent  We will follow the same procedure of calculating radiative
corrections as is done for the
Coulomb potential in QED.  The one loop diagrams are shown in Fig. 1.
The coupling to an
external graviton field $h_{\mu \nu} ^{ext}$ involves the energy momentum
tensor
\begin{equation}
{\cal L}_I = -{\kappa\over 2} h_{\mu \nu}^{ext} T^{\mu \nu}
\end{equation}

\noindent  For an external spinless source with Lagrangian,
\begin{equation}
{\cal L}_M = {\sqrt{-g}\over 2} \left[ g^{\mu \nu} \partial_\mu \phi
\partial_\nu \phi - m^2
\phi^2 \right]
\end{equation}

\noindent  the tensor is
\begin{equation}
T_{\mu \nu}^M =  \partial_\mu \phi \partial_\nu \phi - {1\over 2}
\eta_{\mu \nu} (\partial_\lambda
\phi \partial^\lambda \phi - m^2 \phi^2)
\end{equation}

\noindent  while for two gravitons it is longer
\begin{eqnarray}
T_{\mu \nu}^h = &-& h^{\sigma \lambda} \partial_\mu \partial_\nu
h_{\sigma \lambda} +
{1\over 2} h \partial_\mu \partial_\nu h \nonumber\\
&+& ({1\over 4} \partial_\mu \partial_\nu -
{3\over 8} \eta_{\mu \nu} \Box ) \left[ hh -
2h^{\sigma \lambda} h_{\sigma \lambda} \right] \nonumber\\
&-& \Box \left[ h_{\sigma \mu} h_\nu^\sigma - hh_{\mu \nu}\right]
 \nonumber\\
&-& (\partial_\lambda \partial_\mu \left[ hh^\lambda_\nu \right]
+ \partial_\lambda \partial_\nu
\left[ hh_\mu^\lambda \right]) \nonumber\\
&+& 2 \partial_\sigma \partial_\lambda
\left[ h^\sigma_\mu h^\lambda_\nu - h^{\sigma \lambda}
h_{\mu \nu} - {1\over 2} \eta_{\mu \nu}
h^{\lambda \rho} h_\rho^\sigma + {1\over 2} \eta_{\mu
\nu} hh^{\sigma \lambda} \right] \nonumber\\
&+& 2 \partial_\lambda \left[ h^{\lambda \sigma}
\partial_\mu h_{\sigma \nu} + h^{\lambda \sigma}
\partial_\nu h_{\sigma \mu} \right] \nonumber\\
&-& (h^\sigma_\mu \Box h_{\sigma \nu} + h^\sigma_\nu \Box h_{\sigma \mu}
- h_{\mu \nu} \Box h) \nonumber\\
&+& {\eta_{\mu \nu} \over 2} \left[ h^{\lambda \sigma}
\Box h_{\lambda \sigma} - {1\over 2} h \Box h \right]
\end{eqnarray}

\noindent  where $h = h^\lambda _\lambda$.  The two graviton matter
vertex in Fig. 1b follows
from the Lagrangian
\begin{eqnarray}
{\cal L}_2 = &+& \kappa^2({1\over 2} h^{\mu \lambda} h_\lambda^\nu - {1\over 2}
hh^{\mu \nu})
\partial_\mu
\phi \partial_\nu \phi \nonumber\\
&-& {\kappa^2 \over 8} (h^{\lambda \sigma} h_{\lambda \sigma} - {1\over 2} hh)
\left[ \partial_\mu \phi
\partial^\mu \phi - m^2 \phi^2 \right]
\end{eqnarray}

\noindent  Gauge fixing is accomplished in path integral quantization by
use of Fadeev-Popov ghosts, $\eta_\mu$. The ghost Lagrangian is \cite{3}
\begin{equation}
{\cal L}_{ghost} = \sqrt{-g} \eta^{\mu *} \left[\Box \eta_{\mu \nu}
- R_{\mu \nu} \right] \eta^\nu   .
\end{equation}

\noindent
Collectively, these Lagrangians define the
vertices required to compute Feynman diagrams.

The calculation of the vertex correction is straightforward but algebraically
tedious.  Diagram 1c
does not lead to any non-analytic terms because
the coupling is to the massive particle.  [It does have
an infrared divergence like the one in QED, which can be handled in a
similar fashion
\cite{2}].  In general the radiative corrected matrix element will have the
form
\begin{eqnarray}
V_{\mu \nu} = <p' \mid T_{\mu \nu} \mid p>=& F_1 (q^2) \left[{p'}_\mu p_\nu
+ p_\mu
{p'}_\nu + q^2
\eta_{\mu \nu} \right] \nonumber\\
& + F_2 (q^2) \left[ q_\mu q_\nu - g_{\mu \nu} q^2 \right]
\end{eqnarray}

\noindent  with $F_1 (0) = 1$.  For the first two diagrams the
non-analytic terms are found to be
\begin{eqnarray}
1a: & \Delta F_1 = {\kappa^2 q^2 \over 32\pi^2}  \left[-{3\over 4} ln (-q^2) +
{1\over 16} { \pi^2 m\over \sqrt{-q^2}}
\right] ; &\Delta F_2 = {\kappa^2 m^2 \over 32\pi^2}
\left[ 3 ln(-q^2) + {7\over 8} {\pi^2 m\over \sqrt{-q^2}} \right] \nonumber\\
1b: & \Delta F_1 = 0 ; &
\Delta F_2 =  {\kappa^2 m^2 \over 32\pi^2} \left[- {13\over 3} ln(-q^2) \right]
\nonumber\\
\end{eqnarray}

\noindent  so that

\begin{eqnarray}
F_1(q^2) = & 1 + {\kappa^2 \over 32\pi^2} q^2
\left[ - {3\over 4} ln (-q^2) + {1\over 16} {\pi^2 m\over
\sqrt{-q^2}} \right] + \dots \nonumber\\
F_2 (q^2) = &{\kappa^2 m^2\over 32\pi^2} \left[ -{4 \over 3} ln (-q^2)
+ {7\over 8} {\pi^2 m\over \sqrt(-q^2)} \right]
\end{eqnarray}

The vacuum polarization diagram has been calculated previously \cite{3}. In
dimensional regularization with only massless particles the $ln(-q^2)$ terms
can be read off from the coefficient of the ${1 \over (d-4)}$ pole in a one
loop
graph. This yields
the non-analytic
terms
\begin{equation}
P_{\mu\nu , \alpha\beta} \Pi_{\alpha\beta , \gamma\delta}
P_{\gamma\delta , \rho\sigma} =
{\kappa^2\over
32\pi^2} q^4 \left[ {21\over 120} (\eta_{\mu\rho} \eta_{\nu\sigma} +
\eta_{\mu\sigma}
\eta_{\nu\rho}) + {1\over 120} \eta_{\mu\nu}\eta_{\rho\sigma} \right]
\left[ - ln (-q^2) \right]
\end{equation}

\noindent  where I have dropped many terms proportional to $ q_\mu ,
q_\nu$ etc. which because of gauge invariance do not
contribute to the interaction described below.

The most precise statement of the one loop results is in terms of the
relativistic forms given above,
Eq. 13 - 16.  However, it is pedagogically useful to combine these to define
a potential. I will define this as the sum of one particle reduceable
diagrams. For a two
body interaction, one obtains this potential from the Fourier
transform of the nonrelativistic limit of
Fig. 2, where the blobs indicate the radiative corrections.  In
momentum space we have
\begin{eqnarray}
{-\kappa^2\over 4} {1\over2m_1} V^{(1)}_{\mu\nu} (q)
\left[i D_{\mu\nu, \alpha\beta} (p) +i D_{\mu\nu,
\rho\sigma}i \Pi_{\rho\sigma, \eta\lambda}i D_{\eta\lambda, \alpha\beta}
\right]
V^{(2)}_{\alpha\beta} (q) {1\over 2m_2} \nonumber\\
\approx 4\pi Gm_1 m_2 \left[ {i\over {\bf {q}}^{2}} -
{i\kappa^2\over 32\pi^2} \left[- {127\over 60} ln {\bf {q}}^{2} +
{\pi^2 (m_1 + m_2)\over 2\sqrt{{\bf {q}}^{2}}} \right] \right]
\end{eqnarray}

\noindent  where the second line corresponds to the nonrelativistic limit
$p_\mu = (m, 0), q = (0,
{\bf {q}}~)$.  In taking the Fourier transforms, we use
\begin{eqnarray}
\int {d^3 q\over (2\pi)^3} e^{-i{\bf q\cdot r}} {1\over {\bf q}^2} =
{1\over 4\pi r}
\nonumber\\
\int {d^3 q\over (2\pi)^3} e^{-i{\bf q\cdot r}} {1\over q} = {1\over 2\pi^2
r^2}
\nonumber\\
\int {d^3 q\over (2\pi)^3} e^{-i{\bf q\cdot r}} ln {\bf q}^2 =
{-1\over 2\pi^2 r^3}
\end{eqnarray}

If we reinsert powers of $\hbar$ and c at this stage,
we obtain the potential energy
\begin{equation}
V(r) = -{GM_1 M_2 \over r} \left[ 1 - {G(M_1 + M_2) \over rc^2} -
{127\over 30\pi^2}
{G \hbar\over r^2
c^3}  \right]
\end{equation}

\noindent  The first correction, of order $GM/rc^2$, does not contain
any power of $\hbar$, and is
of the same form as various post-Newtonian corrections which we have
dropped in
taking the
nonrelativistic limit \cite{1}.  In fact, for a small test particle $M_2$,
this piece is the same as the expansion of the time component of the
Schwarzschild metric,
\begin{equation}
g_{00} = {1-{GM_1 \over r c^2} \over 1 + {GM_1  \over r c^2}} \approx
1 - {2GM_1 \over r c^2}\left[ 1 - {GM_1 \over r c^2}\right]
\end{equation}
\noindent which is the origin of the static gravitational potential.
Therefore we do not count this result as a
quantum correction.  However the
last term is a true quantum effect, linear in $\hbar$.  We note also
that if the photon and
neutrinos are truly massless, they too
must be included in the vacuum polarization diagram.  Using the
results of Ref. 8, this changes the quantum modification to
\begin{equation}
  -{135 + 2N_\nu \over 30\pi^2} {G \hbar \over r^2c^3}
\end{equation}

\noindent  where $N_\nu$ is the number of massless helicity states of
neutrinos.

The effect calculated here is distinct from another finite contribution to
the energy momentum vertex - the trace anomaly \cite{14}. The trace
anomaly is a local effect and is represented by analytic corrections to the
vertices, while the crucial distinction is that the non-analytic terms
are non-local.  Note that the quantum correction above
is {\em far} too small to be measured.  However, the specific
number is less important than
the knowledge that a prediction can be made.

The ability to make long distance predictions certainly does not solve
all of the problems of
quantum gravity.  Most likely the theory must be greatly modified at
short distances, for example
as is done in string theory.  Most quantum predictions involving
gravity treat quantum matter fields
in a classical gravitational field \cite{14}.  True predictions
(observable in principle and without unknown parameters) involving the
quantized gravitational field are few.  However,
the methodology of effective field theory, when
applied to gravity, yields well defined quantum predictions at large
distances.

{\bf Acknowledgments}:  I would like to thank
Jennie Traschen and David Kastor for numerous
discussions on this topic and
G. Esposito-Farese, S. Deser, H. Dykstra, E. Golowich, B.
Holstein, G. Leibbrandt and J. Simon for useful comments.

{\bf{Figure Captions}}

\noindent Fig. 1.   One loop radiative corrections to the gravitational vertex
(a-d) and vacuum polarization (e,f).

\noindent Fig. 2.  Diagrams included in the potential. The dots indicate
vertices and propagators including the corrections shown in Feg. 1.

\end{document}